# Mining Top-k Approximate Frequent Patterns[*]

Zengyou He

Department of Computer Science and Engineering,Harbin Institute of Technology,

92 West Dazhi Street, P.O Box 315, Harbin 150001, P. R. China

zengyouhe@yahoo.com

**Abstract** Frequent pattern (itemset) mining in transactional databases is one of the most well-studied problems in data mining. One obstacle that limits the practical usage of frequent pattern mining is the extremely large number of patterns generated. Such a large size of the output collection makes it difficult for users to understand and use in practice. Even restricting the output to the border of the frequent itemset collection does not help much in alleviating the problem.

In this paper we address the issue of overwhelmingly large output size by introducing and studying the following problem: *mining top-k approximate frequent patterns*. The union of the power sets of these *k* sets should satisfy the following conditions: (1) including itemsets with larger support as many as possible and (2) including itemsets with smaller support as less as possible. An integrated objective function is designed to combine these two objectives. Consequently, we derive the upper bounds on objective function and present an approximate branch-and-bound method for finding the feasible solution. We give empirical evidence showing that our formulation and approximation methods work well in practice.

## 1. Introduction

Frequent pattern mining is a widely studied topic in data mining research [e.g., 1-3]. Its popularity can be attributed to the simplicity of the problem statement and the efficiency of support-based pruning strategy to prune a combinatorial space.

However, end users of traditional frequent pattern mining applications encounter several well-known problems in practice. First, without specific knowledge about the target data, users will have difficulties in setting the support threshold to obtain their required results. Second, the algorithms often generate an extremely large number of frequent patterns, often in thousands or millions, which is even larger than the original target dataset. It is nearly impossible for the end users to comprehend or validate such large number of complex patterns, thereby limiting frequent pattern mining spread use and acceptance in many real-world situations.

As pointed out in [4], the collection of frequent patterns can be used in at least two different ways: first, one can be interested in the individual patterns and their occurrence; second, one can be interested in the whole collection, trying to obtain a global view of which patterns are frequent and which are not. However, in many cases the collection of frequent patterns is large, and obtaining a global understanding of which patterns are frequent and which are not is not easy.

[*] This work was supported by the High Technology Research and Development Program of China (No. 2003AA4Z2170, No. 2003AA413021) and the IBM SUR Research Fund.

Even restricting the output to the border of the frequent itemset collection does not help much in alleviating the problem [4].

From the second global view, the purpose of this paper is to address the issue of overwhelmingly large output size by introducing and studying the following problem: *mining top-k approximate frequent patterns*. We aim at finding only *k* set that is easy-to-understand and approximate those most frequent patterns as well as possible. The premise of our work is that such small approximations provide a better understanding on the global structure of the data set without a significant sacrifice of information. Indeed, whether a pattern is frequent is highly correlated with the support threshold, and this threshold is almost always somewhat arbitrary. Hence, reporting only a small collection of itemsets for approximating most frequent patterns is more desirable in practice.

In general, we are interested in longer itemset since then it could cover more patterns to provide better summarization. We are also interested in itemset that contains a large portion of most frequent patterns. A natural way to combine these objectives is to maximize their product so that we simultaneously ensure it contains many top frequent patterns and also that many lower-support patterns are excluded.

To efficiently solve the problem of top-*k* approximate frequent patterns, we first derive the upper bounds of objective function, and then present an approximate branch-and-bound method for finding the feasible solution. We give empirical evidence showing that our formulation and approximation methods work well in practice.

This paper is organized as follows: Section 2 discusses the related work. Section 3 defines the problem as an optimization problem in detail. Then the optimistic bounds on objective functions are provided and proved in Section 4. In Section 5, a branch-and-bound method is presented to find best *k* sets. Finally, Section 6 gives empirical results and Section 7 concludes the paper.

## 2. Related Work

Recently, we have witnessed an increasing interest in devising new non-traditional methods that produce more manageable chunks of knowledge rather than high-volume frequent patterns. In general, these efforts can be summarized as the following categories.

Firstly, a number of more concise representations of frequent itemsets have been proposed, such as frequent closed itemsets [5], the generators representation [6], maximal frequent itemset [7,8], etc. The work on frequent closed item sets attempts to compress the collection of frequent sets in a lossless manner, while for the maximal frequent itemset the output is restricted to the border of the frequent itemset collection.

Secondly, recent work has highlighted the importance of constraint-based mining paradigm in the context of frequent patterns [e.g., 9]. Some categories of constraints considered so far include item constraint, length constraint, model-based constraint and aggregate constraint. It is expected to reduce the overwhelming number of patterns by imposing additional constraints on the pattern discovery process.

Thirdly, some researchers have begun to consider the task of mining *k* most frequent (closed) patterns instead of mining frequent patterns [10-13]. Shen et al. [10] considers the problem of mining *N* largest itemsets. Cheung and Fu [11] proposed a different formulation by mining *N*

$k$-itemsets with the highest supports for $k$ up to a certain $k_{max}$ value. Han et al. [12] proposed another new mining task: mining top-$k$ frequent closed patterns of length no less than *min _l*, where $k$ is the desired number of frequent closed patterns to be mined, and *min _l* is the minimal length of each pattern. As an extension to [12], the problem of mining top-$k$ sequential closed patterns of length no less than *min _ l* is discussed in [13].

Finally, a post-processing technique by finding a succinct representation of a collection of frequent sets is proposed in [4], which is most related with our work. However, there are significant differences between our method and the work in [4] with respect to both problem formulation and underlying methodology. In the problem of [4], the input is a collection of frequent sets, while our input is the original transaction database so that the costly computation for set of frequent patterns is avoided. Indeed, a post-processing methodology is adopted in [4] and hence it is still sensitive to support threshold, while a new kind of "knowledge or data mining task is defined in this paper since then it can be directly derived from the original databases.

## 3. Problem Formulation

Agrawal's statement of the problem of discovering frequent itemset in market basket databases is the following [1].

Let $I = \{i_1, i_2, \ldots, i_m\}$ be a set of $m$ literals called *items*. Let the database $D = \{T_1, T_2, \ldots, T_n\}$ be a set of $n$ transactions each consisting of a set of items from $I$. An itemset $X$ is a non-empty subset of $I$. The length of itemset $X$ is the number of items contained in $X$, and $X$ is called a *k-itemset* if its length is $k$. A transaction $t \in D$ is said to contain an itemset $X$ if $X \subseteq t$. The *support* of an itemset $X$ is the number of transactions in $D$ containing $X$: $\sigma(X) = |\{t \in D \mid X \subseteq t\}|$, where $|\{.\}|$ denotes the number of elements that belong to a given set. An itemset is *frequent* if $\sigma(X) \geq minsup$, where *minsup* is a specified minimum support threshold. The power set of $X$ is denoted as *Pow* ($X$).

The problem of finding all *frequent itemsets* in $D$ is then traditionally defined as follows. Given user defined support threshold *minsup*, find all itemsets with support greater or equal to *minsup*. Frequent itemsets are also called *frequent patterns*.

Before formally introducing the problem of mining top-$k$ approximate frequent patterns, we first design a reasonable objective function for evaluating the goodness of each candidate. To clarify our idea, we first consider the special case when $k=1$.

To make a pattern $P$ covers most frequent patterns as many as possible, it is very natural to prefer longer pattern (larger $|P|$). To avoid overgeneralization, we also hope that only those most frequent patterns are included in the power set of $P$. Hence, the average support of the patterns in *Pow* ($P$) is used for this purpose, i.e., $\dfrac{\sum_{SP \in Pow(P)} \sigma(SP)}{2^{|P|}-1}$. A natural way to combine these objectives is to maximize $|P| \cdot \dfrac{\sum_{SP \in Pow(P)} \sigma(SP)}{(2^{|P|}-1)}$ so that we simultaneously ensure it contains many top frequent patterns and also that many lower-support patterns are excluded.

Computing the value of $|P| \cdot \dfrac{\sum_{SP \in Pow(P)} \sigma(SP)}{(2^{|P|} - 1)}$ needs to know the support of each subset of $P$. Such requirements will encounter costly computation. To reduce the computation cost, the object function is transformed into $|P| \cdot \dfrac{\sum_{T \in D}(2^{|T \cap P|} - 1)}{(2^{|P|} - 1)}$ since the equation $\sum_{SP \in Pow(P)} \sigma(SP) = \sum_{T \in D}(2^{|T \cap P|} - 1)$ holds, where each $|T \cap P|$ is much easier to compute from the database. Therefore, the problem of mining top-1 approximate frequent pattern could be stated as follows (Problem 1):

**Problem 1.** Given the database $D = \{T_1, T_2, \ldots, T_n\}$ with $n$ transactions and each transaction consisting of a set of items from $I=\{i_1, i_2, \ldots, i_m\}$, find a pattern $P \in I$ such that formula (1) is maximized.

$$|P| \cdot \dfrac{\sum_{T \in D}(2^{|T \cap P|} - 1)}{(2^{|P|} - 1)} \qquad (1)$$

We also call this problem as the problem of best approximate frequent pattern mining. The optimal solution to this problem for all cases is denoted by $P^*$.

Naturally, the problem of mining top-$k$ approximate frequent pattern could be formulated as follows (Problem 2)

**Problem 2.** Given transaction database $D$ and a positive integer $k$, find $k$ patterns with largest object value (formula (1)).

In problem 2, we are aiming at find $k$ sets that best approximate the most frequent patterns, in which the object function is used as the same with Problem 1.

## 4. Upper Bounds for Pruning Search Space

Knowing the upper bounds of the object function will help us prune the exponential search space and find the feasible solution in a more efficient way.

**Lemma 1:** For $1 \leq i \leq j$ ($i$ and $j$ are integers), $\dfrac{2^i}{i} \leq \dfrac{2^j}{j}$ holds.

**Proof:** First, the continuous function $f(x) = \dfrac{2^x}{x}$ is increasing on $[2, +\infty)$ because the derivative $f'(x) = \dfrac{(ln2)x - 1}{x} > 0$ when $x \geq 2$. That is, for any $2 \leq i \leq j$, $\dfrac{2^i}{i} \leq \dfrac{2^j}{j}$ holds.

Moreover, $\dfrac{2^1}{1} \leq \dfrac{2^2}{2}$ holds. Hence, for $1 \leq i \leq j$ ($i$ and $j$ are integers), $\dfrac{2^i}{i} \leq \dfrac{2^j}{j}$ holds. □

**Lemma 2:** For positive integer $x$, if $t \geq 3$, then $(1+x/t)^2 \leq 2^x$ holds.

**Proof:** $(1+x/t)^2 \leq 2^x \Leftrightarrow t \geq \dfrac{x}{2^{x/2}-1}$. $f(x) = \dfrac{x}{2^{x/2}-1}$ is a decreasing function on $[1,+\infty)$ because the derivative $f'(x) = \dfrac{2^{x/2}(1-(In2)x/2)-1}{(2^{x/2}-1)^2} < 0$ when $x \geq 1$. Furthermore, $f(1) = \dfrac{1}{2^{1/2}-1} = \sqrt{2}+1$. Hence, when $t \geq 3$, we have $t > f(1) \geq f(x) = \dfrac{x}{2^{x/2}-1}$. That is, we can draw a conclusion that if $t \geq 3$, then $(1+x/t)^2 \leq 2^x$ holds. □

**Theorem 1:** Given the transaction database $D$, the upper bound of $|P| \cdot \dfrac{\sum\limits_{T \in D}(2^{|T \cap P|}-1)}{(2^{|P|}-1)}$ is $\dfrac{|P| \cdot 2^q}{q \cdot (2^{|P|}-1)} \sum\limits_{T \in D} |T \cap P|$, where $q = min\{|P|, max\{|T| \, | \, T \in D\}\}$.

**Proof:** Firstly, we know that $0 \leq |T \cap P| \leq q$ and $|T \cap P|$ is an integer, so $\sum\limits_{T \in D} 2^{|T \cap P|}$ can be re-written as: $\sum\limits_{i=0}^{q} c_i \cdot 2^i$, where $c_i$ is times that $|T \cap P| = i$. Then, $\sum\limits_{i=1}^{q} c_i \cdot i = \sum\limits_{T \in D} |T \cap P|$ holds. Hence, we have

$$\sum_{i=0}^{q} c_i \cdot 2^i = c_0 \cdot 2^0 + \sum_{i=1}^{q} c_i \cdot 2^i = c_0 + \sum_{i=1}^{q} c_i \cdot 2^i \quad // \ c_0 \text{ is times that } |T \cap P| = \phi$$

$$\leq (n - min\{\sigma(S) \, | \, S \in P\}) + \sum_{i=1}^{q} c_i \cdot 2^i = (n - min\{\sigma(S) \, | \, S \in P\}) + \sum_{i=1}^{q} c_i \cdot i \cdot (\dfrac{2^i}{i})$$

$$\leq (n - min\{\sigma(S) \, | \, S \in P\}) + \sum_{i=1}^{q} c_i \cdot i \cdot (\dfrac{2^q}{q}) \quad // \text{ Lemma 1}$$

$$= (n - min\{\sigma(S) \, | \, S \in P\}) + \dfrac{2^q}{q} \cdot \sum_{i=1}^{q} c_i \cdot i$$

$$= (n - min\{\sigma(S) \, | \, S \in P\}) + \dfrac{2^q}{q} \cdot \sum_{T \in D} |T \cap P|$$

Therefore, $|P| \cdot \dfrac{\sum\limits_{T \in D}(2^{|T \cap P|}-1)}{(2^{|P|}-1)} = \dfrac{|P| \cdot \sum\limits_{T \in D} 2^{|T \cap P|}}{(2^{|P|}-1)} - \dfrac{|P| \cdot n}{(2^{|P|}-1)}$

$$\leq \frac{|P| \cdot (n - \min\{\sigma(S) \mid S \in P\} + \frac{2^q}{q} \cdot \sum_{T \in D} |T \cap P|)}{(2^{|P|} - 1)} - \frac{|P| \cdot n}{(2^{|P|} - 1)}$$

$$= \frac{|P| \cdot (\frac{2^q}{q} \cdot \sum_{T \in D} |T \cap P| - \min\{\sigma(S) \mid S \in P\})}{(2^{|P|} - 1)}$$

$$= \frac{|P| \cdot 2^q}{q \cdot (2^{|P|} - 1)} \sum_{T \in D} |T \cap P| - \frac{|P| \cdot \min\{\sigma(S) \mid S \in P\}}{2^{|P|} - 1}$$

$$< \frac{|P| \cdot 2^q}{q \cdot (2^{|P|} - 1)} \sum_{T \in D} |T \cap P| \qquad \square$$

Theorem 1 presents a lower bound that can be computed using only the support of each single item since $\sum_{T \in D} |T \cap P| = \sum_{S \in P} \sigma(S)$. It means that we can get the lower bound of any pattern $P$ without accessing the database.

From Theorem 1, we also know that if a pattern contains items with larger support, it has larger upper bound. In order to efficiently find a feasible solution with a branch-and-bound strategy (we will discuss the algorithm in detail in next section), the items are organized in a decreasing order with respect to their support values, i.e., $\forall i_u, i_v \in I = \{i_1, i_2, \ldots, i_m\}$, if $u<v$, then $\delta(i_u) \geq \delta(i_v)$. We use $pos(i_j)$ denoted the position of item $i_j$ in $I$, i.e., $pos(i_j)=j$. And the candidate pattern is generated and searched in a depth-first manner. In such situation, we have Theorem 2 holds.

**Theorem 2:** Given the transaction database $D$, when $|P| \geq q \geq 3$, the objective value of any pattern $Q$ with $P$ as its prefix has an upper bound of $\frac{|P| \cdot 2^q}{q \cdot (2^{|P|} - 1)} \sum_{T \in D} |T \cap P|$, where $q = \max\{|T| \mid T \in D\}$.

**Proof:** From Theorem 1, we know that $\frac{|P| \cdot 2^q}{q \cdot (2^{|P|} - 1)} \sum_{T \in D} |T \cap P|$ is upper bound on the objective value of pattern $P$. Hence, to prove Theorem 2, we only need to show that: when $|P| \geq q \geq 3$, $\frac{|P| \cdot 2^q}{q \cdot (2^{|P|} - 1)} \sum_{T \in D} |T \cap P| \geq \frac{|Q| \cdot 2^q}{q \cdot (2^{|Q|} - 1)} \sum_{T \in D} |T \cap Q|$ holds.

Since $Q$ take $P$ as its prefix, we can assume that $Q = P \cup R$, where $P \cap R = \Phi$ and all items in $R$ are after the items in $P$ in the ordered list, i.e., $\forall x \in P, \forall y \in R, \delta(x) \geq \delta(y)$.

To prove $\dfrac{|P| \cdot 2^q}{q \cdot (2^{|P|}-1)} \sum_{T \in D} |T \cap P| \geq \dfrac{|Q| \cdot 2^q}{q \cdot (2^{|Q|}-1)} \sum_{T \in D} |T \cap Q|$, we only need to show that

$$\dfrac{|P|}{(2^{|P|}-1)} \sum_{SP \in P} \delta(SP) \geq \dfrac{|Q|}{(2^{|Q|}-1)} \sum_{SQ \in Q} \delta(SQ) = \dfrac{|P|+|R|}{(2^{|P|} \cdot 2^{|R|}-1)} (\sum_{SP \in P} \delta(SP) + \sum_{SR \in R} \delta(SR)) \text{ holds.}$$

Moreover, we have

$$\dfrac{|P|+|R|}{(2^{|P|} \cdot 2^{|R|}-1)} (\sum_{SP \in P} \delta(SP) + \sum_{SR \in R} \delta(SR)) < \dfrac{|P|+|R|}{2^{|R|}(2^{|P|}-1)} (\sum_{SP \in P} \delta(SP) + \sum_{SR \in R} \delta(SR))$$

$$\leq \dfrac{|P|+|R|}{2^{|R|}(2^{|P|}-1)} (\sum_{SP \in P} \delta(SP) + (|R|/|P|) \sum_{SP \in P} \delta(SP)) \quad // \forall x \in P, \forall y \in R, \delta(x) \geq \delta(y)$$

$$= \dfrac{(|P|+|R|)^2}{2^{|R|}(2^{|P|}-1)|P|} \sum_{SP \in P} \delta(SP)$$

Hence, it is equivalent to show that $\dfrac{|P|}{(2^{|P|}-1)} \sum_{SP \in P} \delta(SP) \geq \dfrac{(|P|+|R|)^2}{2^{|R|}(2^{|P|}-1)|P|} \sum_{SP \in P} \delta(SP)$, i.e.,

$2^R \geq (1+|R|/|P|)^2$. According to Lemma 2, $2^R \geq (1+|R|/|P|)^2$ holds when $|P| \geq 3$. Therefore, when $|P| \geq q \geq 3$, the objective value of any pattern $Q$ with $P$ as its prefix has an upper bound of $\dfrac{|P| \cdot 2^q}{q \cdot (2^{|P|}-1)} \sum_{T \in D} |T \cap P|$. □

Theorem 2 provides us an upper bound on the objective values of all the patterns with $P$ as prefix. Hence, in branch-and-bound search procedure, we can just neglect branches generated from $P$ when this upper bound is smaller than current best objective value already obtained. As we can see, $q \geq 3$ is required to ensure the validity of the general upper bound, which is almost satisfied by all transaction database since the longest item usually has length greater than 2.

In Theorem 2, the upper bound is valid only when $|P| \geq q$ holds. To get an upper bound for patterns with any pattern $P$ as prefix with arbitrary length, we have to consider the upper bound when $|P| \leq q$, as stated in Theorem 3.

**Theorem 3:** Given the transaction database $D$ with $I = \{i_1, i_2, …, i_m\}$ ordered by support values, when $|P| \leq q$, the objective value of any pattern $Q$ with $P$ as its prefix has an upper bound $UB(Q)$ defined as follows (suppose that the item in $P$ with largest position is denoted as $i_c$):

$$UB(Q) = \max_{j=pos(i_c)}^{pos(i_c)+\min\{m-pos(i_c),q-|P|\}} \left\{ \frac{|P \cup (\bigcup_{h=pos(i_c)}^{j} i_h)|}{(2^{|P \cup (\bigcup_{h=pos(i_c)}^{j} i_h)|} - 1)} \sum_{T \in D} |T \cap (P \cup (\bigcup_{h=pos(i_c)}^{j} i_h))| \right\} \quad (2)$$

**Proof:** According to whether $|Q| > q$ or not, we have two cases.

**CASE 1:** $|Q| \leq q$. Since the items in $I$ are in a decreasing order with respect to support values, hence, the upper bound of $Q$ is not greater than the upper bound of pattern $P \cup (\bigcup_{h=pos(i_c)}^{h+|Q|-|P|} i_h)$.

Therefore, in this case, the upper bound $Q$ is less than the $UB(Q)$ defined in (2).

**CASE 2:** $|Q| > q$. In this case, we know that $m - pos(i_c) \geq q - |P|$ hold.

If $Q$ has the pattern $P \cup (\bigcup_{h=pos(i_c)}^{h+q-|P|} i_h)$ as its prefix, then according to Theorem 2, the upper bound of $Q$ is not greater than the upper bound of pattern $P \cup (\bigcup_{h=pos(i_c)}^{h+q-|P|} i_h)$.

Otherwise, we can construct a pattern $P \cup (\bigcup_{h=pos(i_c)}^{h+|Q|-|P|} i_h)$, whose upper bound is not greater than the upper bound of $P \cup (\bigcup_{h=pos(i_c)}^{h+q-|P|} i_h)$ according to Theorem 2. Furthermore, consider again the fact that the items in $I$ are in a decreasing order with respect to support values, we know the upper bound of $Q$ is less than $P \cup (\bigcup_{h=pos(i_c)}^{h+q-|P|} i_h)$. Hence, the upper bound $Q$ is less than the $UB(Q)$ defined in (2).

Considering both Case 1 and Case 2, the proof is finished. □

From Theorem 2 and Theorem 3, we can straightforwardly get a general upper bound for patterns with any pattern $P$ as prefix with arbitrary length, as summarized in Theorem 4.

**Theorem 4:** Given the transaction database $D$ with $I = \{i_1, i_2, \ldots, i_m\}$ ordered by support values, when $q = max\{|T| \,|\, T \in D\} \geq 3$, the objective value of any pattern $Q$ with $P$ as its prefix has an upper bound $UB(Q)$ defined as follows (suppose that the item in $P$ with largest position is denoted as $i_c$):

$$UB(Q) = \begin{cases} \dfrac{|P| \cdot 2^q}{q \cdot (2^{|P|} - 1)} \sum_{T \in D} |T \cap P|, & \text{if } |P| \geq q \\ \max_{j=pos(i_c)}^{pos(i_c)+\min\{m-pos(i_c), q-|P|\}} \left\{ \dfrac{|P \cup (\bigcup_{h=pos(i_c)}^{j} i_h)|}{2^{|P \cup (\bigcup_{h=pos(i_c)}^{j} i_h)|} - 1} \sum_{T \in D} |T \cap (P \cup (\bigcup_{h=pos(i_c)}^{j} i_h))| \right\}, & \text{if } |P| \leq q \end{cases}$$

## 5. ABB: An Approximate Branch-and-Bound Search Method

### 5.1 Best Approximate Frequent Pattern Mining

The branch and bound method is a classical technique in combinatorial optimization. It uses an ordered search method on a partitioning of the search spaces. In our problem, finding good upper bound on the target search spaces does pruning. If this upper bound is less than the some solution that has been determined, then the corresponding candidates may be pruned for contention.

As have been discussed in Section 4, the items are organized in a decreasing order with respect to their support values. Then, our algorithms will follow a similar depth-first search strategy as used in the Eclat algorithm [2]. The same search strategy was later also successfully used in the FP-growth algorithm [3], and is based on a divide and conquer mechanism. In the searching process, we first get the upper bound on the objective values of all patterns with current candidate as prefix. If the upper bound is less than the best solution that has been determined, this branch could be pruned without further considerations.

The above branch and bound method is able to find optimal solution. However, the potential search spaces are extremely large, even after the use of upper bound based pruning. To tackle this problem, we proposed an approximate branch and bound method for finding feasible solutions more efficiently.

In the approximate branch and bound method, we would introduce three additional parameters: approximation ratio (denoted by *ar*), *epoch* and *delta*. Whenever *epoch* additional branches are generated and visited, the approximation ratio *ar* is increased by *delta*. In the searching process, we first get the upper bound of all patterns with current candidate as prefix. If the upper bound divided by *ar*, is less than the best solution that has been determined, this branch is pruned.

The approximate branch and bound algorithm ABB as shown in Fig.1, recursively transverse all branches in a depth-first manner.

We now describe the ABB algorithm in more detail. First, the algorithm is initialized, where the prefix pattern is set to be empty and best objective value is set to 0. Then, in Step (02), we get the upper bound of all patterns than begin with prefix. If this value is greater than the product of current best objective value and the approximation ratio, we proceed further; otherwise, no further search is needed on branches with this prefix. The computation of upper bound is efficient since we only need to know the support values of single items, which could be derived directly since we utilize the vertical database layout schema as used in Eclat algorithm [2].

In Step (03) and (04), we compute the exact objective value of the prefix. If this value is better than current best objective value, an update operation is executed.

In Step (06)-Step (07), for those items after the items in the prefix, more branches are further generated and ABB (Prefix) is recursively called

Finally, from Step (09)-Step (13), the approximation ratio *ar* is increased by *delta* if *epoch* additional branches are already generated and visited.

```
Algorithm ABB (Prefix)

Input:   D         // the transaction database
         I         // the set of all items
         ar        // initial approximation ratio
         epoch     // an integer
         delta     // amount increased by approximation ratio in each epoch
Output:  An Approximate Frequent Pattern
Initialized
Prefix = {};                 // the initial prefix pattern is set to be empty
Best_Objective_Value = 0;
Number_Of_Branches_Visited =0;

01  Begin
02  If getUpperBound (Prefix)> Best_Objective_Value* ar then
03      If getObjectiveValue (Prefix) > Best_Objective_Value then
04          Best_Objective_Value = getObjectiveValue (Prefix)
05      End If
06      For each item i in I that is after the items in Prefix
07          ABB (Prefix+{i})
08      End For
09      Number_Of_Branches_Visited++;
10      If Number_Of_Branches_Visited> epoch Then
11          ar = ar + delta
12          Number_Of_Branches_Visited = Number_Of_Branches_Visited – epoch
13      End If
14  End If
15  End
```

**Fig. 1** The ABB Algorithm

Obviously, after the execution of the algorithm, we get an approximate solution with approximation bound (1/*ar**), where the *ar** is the value of *ar* when the algorithm terminates.

**5.2 Top-*K* Approximate Frequent Pattern Mining**

In order to find the top-k approximate frequent patterns, we adapt the ABB algorithm as follows. Initially the objective value threshold also is zero. Then, after the algorithm has generated the first *k* patterns with large objective values, it increases objective value threshold to the objective value of the smallest of these *k* patterns. From here on, the threshold can be increased every time a larger pattern is discovered. Then, similarly, the algorithm executes in a recursive manner.

## 6. Experimental Results

We implemented our algorithms in Java and experimented on a 2.4GHz Pentium-4 PC with 512MB of memory, running Windows 2000 Professional.

We present the evaluation of the ABB algorithm for finding best approximate frequent pattern on 4 real datasets from UCI repository of machine learning databases [14], which are the Congressional Votes dataset, the Wisconsin Breast Cancer dataset, the Mushroom dataset and the Zoo dataset.

For evaluating the discovered pattern, we define an additional *coverage* measure besides the objective value. For a pattern *P*, the size of its power set is $2^{|P|}-1$ (denoted by Pow (P)), i.e., $2^{|P|}-1$ itemsets could be derived from it. If there are more elements in intersection between the power set of *P* and the set of top ($2^{|P|}-1$) frequent patterns (denoted by TKP), the pattern *P* is more desired. Hence, *coverage* measure could be defined as: $|Pow(P) \cap TKP|/(2^{|P|}-1)$.

In the first experiments, we test performance of ABB algorithm on the four datasets, as shown in Table 1, where initial approximation ratio is set to 1, *epoch* =1000 and *delta*=0.1.

Table 1 Experimental Results on 4 Datasets

| Measures<br>Dataset | Coverage | Objective Value | Final Approximation Ratio (*ar\**) | Execution Time (sec) |
|---|---|---|---|---|
| Votes | 19.4% | 651.6 | 6.1 | 81.7 |
| Cancer | 99.6% | 2293.1 | 1.4 | 17.4 |
| Mushroom | 100% | 35187.7 | 3.1 | 1070.6 |
| Zoo | 51% | 380.6 | 3.0 | 19.6 |

From Table 1, some important observations are summarized as follows.

(1) For different datasets, the *coverage* may very significantly. Further explorations on these datasets, reveals the fact that, datasets with higher coverage (e.g., mushroom) is much more dense and contain longer frequent patterns. Anyway, our algorithm is able to find a pattern to approximate those most frequent patterns in a satisfactory manner.

(2) Although the final approximation ratio is also different in different datasets, all of them provide us a reasonable approximation bound, which is sufficed for real applications.

In our second experiments, we will examine how the algorithm's output is affected by the input parameter *epoch*. To that goal, on all datasets we range *epoch* from 200 to 1000 to observe the changes on objective values, final approximation ratio and execution times, as shown in Fig.2, Fig.3 and Fig.4 separately.

From those Figures, we found that objective values and final approximation ratio keep stable when the *epoch* is increased, while execution time increases linearly. It indicates that we can set a relatively small *epoch* value to get a feasible solution more efficiently.

For the input parameter *delta*, we observed a similar phenomenon. Therefore, the details are not presented here.

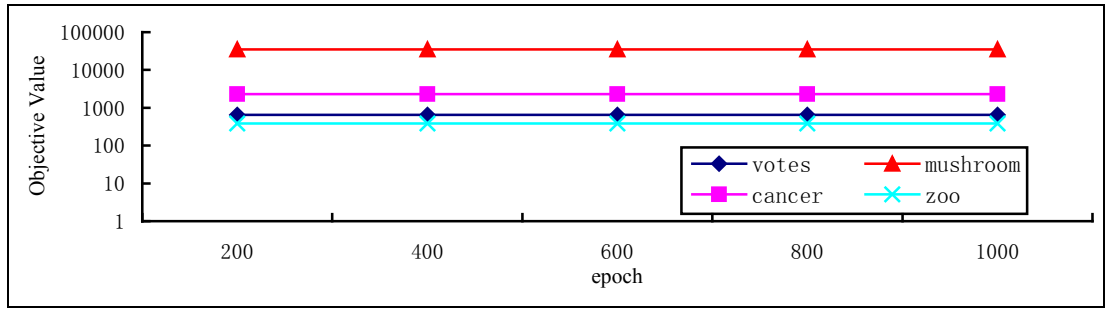

**Fig. 2** The Objective Values vs. Epoch

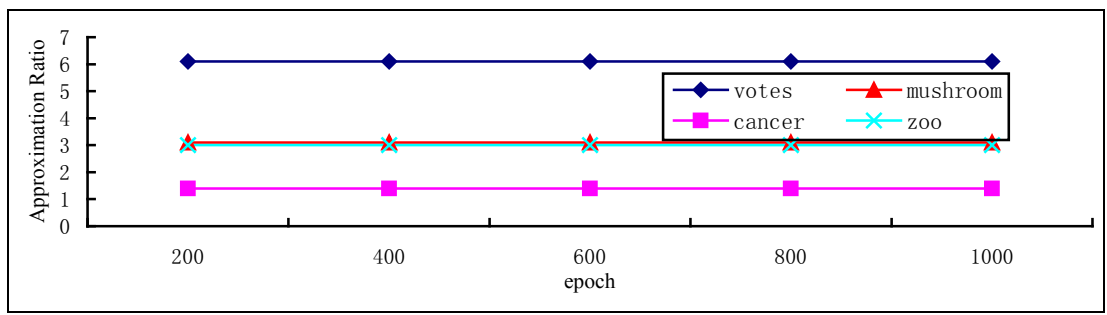

**Fig. 3** The Approximation Ratio vs. Epoch

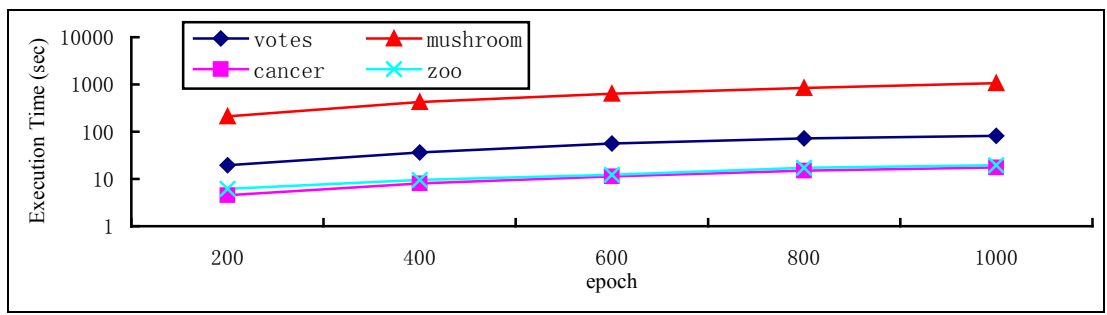

**Fig. 4** The Execution Time vs. Epoch

## 7. Conclusions

We have considered the problem of *mining top-k approximate frequent patterns*. We designed an objective function for evaluating the goodness of each pattern. Consequently, we derive a general upper bound on objective function and present an approximate branch-and-bound method for finding the feasible solution. We provide empirical evidence showing that our formulation and algorithm work well in practice.

# References

[1] R. Agrawal, R. Srikant. Fast Algorithms for Mining Association Rules. In: Proc of VLDB'94, pp. 478-499, 1994.

[2] M. J. Zaki. Scalable Algorithms for Association Mining. IEEE Trans. On Knowledge and Data Engineering, 2000, 12(3): 372~390.

[3] J. Han, J. Pei, J. Yin. Mining Frequent Patterns without Candidate Generation. In: Proc. of SIGMOD'00, pp. 1-12, 2000.

[4] F. N. Afrati, A. Gionis, H. Mannila. Approximating a collection of frequent sets. In: Proc. of KDD'04, pp. 12-19, 2004

[5] N. Pasquier, Y. Bastide, R. Taouil, L. Lakhal. Discovering Frequent Closed Itemsets for Association Rules. In: Proc. of ICDT'99, pp. 398–416,1999.

[6] M. Kryszkiewicz. Concise Representation of Frequent Patterns based on Disjunction–free Generators. In: Proc. of ICDM'01, pp. 305–312,2001.

[7] D-I. Lin, Z. M. Kedem. Pincer Search: A New Algorithm for Discovering the Maximum Frequent Set. In: Proc. of EDBT'98, pp.105-119, 1998.

[8] D. Burdick, M. Calimlim, J. Gehrke. MAFIA: A Maximal Frequent Itemset Algorithm for Transactional Databases. In: Proc .of ICDE'01, pp.443-452, 2001

[9] J. Pei, J. Han. Constrained frequent pattern mining: a pattern-growth view. SIGKDD Explorations, 4 (1): 31-39, 2002.

[10] L. Shen, H. Shen, P. Prithard, R. Topor. Finding the N Largest Itemsets. In Data Mining (Editor: N.F.F. EBECKEN, serve as Proc. of ICDM'98), Great Britain: WIT Pr, 1998.

[11] Y-L.Cheung A. Fu. Mining Frequent Itemsets without Support Threshold: With and without Item Constraints. IEEE Transactions on Knowledge and Data Engineering, 16(9): 1052- 1069, 2004.

[12] J. Han, J. Wang, Y. Lu, P. Tzvetkov: Mining Top-K Frequent Closed Patterns without Minimum Support. In: Proc. of ICDM'02, pp.211-218, 2002.

[13] P. Tzvetkov, X. Yan, J. Han. TSP: Mining Top-K Closed Sequential Patterns. In: Proc. of ICDM'03, pp.347-354, 2003.

[14] C. J. Merz, P. Merphy. UCI Repository of Machine Learning Databases, 1996. ( Http://www.ics.uci.edu/~mlearn/MLRRepository.html).